\documentclass[doublecol]{epl2}

\title{Universality in the emergence of oscillatory instabilities in turbulent flows}

\author{Induja Pavithran\inst{1} \and Vishnu R. Unni\inst{2} \and Alan J. Varghese\inst{3} \and R. I. Sujith\inst{3} \and Abhishek Saha\inst{2} \and Norbert Marwan\inst{4} \and J\"urgen Kurths\inst{4,5,6}}
\shortauthor{I. Pavithran \etal}

\institute{                    
  \inst{1} Department of Physics, IIT Madras, Chennai-600036, India\\
  \inst{2} Department of Mechanical and Aerospace Engineering, University of California, San Diego, USA\\
  \inst{3} Department of Aerospace Engineering, IIT Madras, Chennai-600036, India\\
  \inst{4} Potsdam Institute for Climate Impact Research, Germany\\
  \inst{5} Department of Physics, Humboldt University, Germany\\
  \inst{6} Institute for Complex Systems and Mathematical Biology, University of Aberdeen, United Kingdom
}

\abstract{ Spontaneous emergence of periodic oscillations due to self-organization is ubiquitous in turbulent flows. The emergence of such oscillatory instabilities in turbulent fluid mechanical systems is often studied in different system-specific frameworks. We uncover the existence of a universal scaling behaviour during self-organization in turbulent flows leading to oscillatory instability. Our experiments show that the spectral amplitude of the dominant mode of oscillations scales inversely with the Hurst exponent of a fluctuating state variable following an inverse power law relation. Interestingly, we observe the same power law behaviour with a constant exponent near -2 across various turbulent systems such as aeroacoustic, thermoacoustic and aeroelastic systems.}

\begin{document}

\maketitle

\section{Introduction}
A large number of physical systems involve turbulent flows that have chaotic variations in properties such as pressure and velocity. Turbulent flows are characterized by eddies of different length and time scales that interact nonlinearly. The transfer of energy across eddies of different length scales takes place through various cascade processes \cite{richardson1926atmospheric, kraichnan1967inertial}. A unique collective behaviour can often arise from the interaction of multiple subsystems resulting in various phenomena at many different scales. Turbulent flow systems can therefore be regarded as a complex system. Although turbulent flows are chaotic, self-organization due to feedback in such a complex system can cause the emergence of order from chaos.

Self-organization is a fundamental property of a complex system, where some form of macroscopic order emerges from interactions between subsystems of an initially disordered system. In turbulent flows, spatially extended patterns such as large coherent structures are formed due to self-organization, for example, devastating cyclones in atmospheric flows. Self-organization driven by feedback between subsystems in turbulent systems can lead to oscillatory instabilities as observed in thermoacoustic \cite{juniper2018sensitivity}, aeroacoustic \cite{flandro2003aeroacoustic}, and aeroelastic systems \cite{hansen2007aeroelastic}. These oscillatory instabilities cause high amplitude vibrations which may incur catastrophic effects in engineering systems. In the present work, we study the emergence of such oscillatory instabilities in three different fluid mechanical systems, namely thermoacoustic, aeroacoustic, and aeroelastic systems.

Feedback between turbulent flow and other subsystems is often the cause for oscillatory instabilities. Thermoacoustic instability, a state of self-sustained large amplitude periodic oscillations in the state variables, arises due to the nonlinear coupling between the reactive flow field and the acoustic field in a confinement \cite{lieuwen2005combustion}. This phenomenon can cause structural damages due to the increased thermal and vibrational loads, forcing shutdown of gas turbine engines \cite{lieuwen2005combustion, fleming1998turbine}, or failure of rockets \cite{fisher2009remembering}. Similarly, aeroacoustic instability is caused by the interaction between the acoustic field in a confinement and vortex shedding in turbulent flows \cite{flandro2003aeroacoustic}. Examples include the pleasant sounds generated in a flute or the destructive large amplitude oscillations established in gas-transport pipelines \cite{kriesels1995high}. Aeroelastic instability occurs as a consequence of the interaction of the flow with the structural elements of the system \cite{hansen2007aeroelastic}, \textit{e.g.}, the catastrophic collapse of the Tacoma Bridge \cite{larsen1997aeroelastic}. The transition to such oscillatory instabilities from a state of chaotic oscillations in turbulent systems occurs via intermittency \cite{nair2014intermittency, nair2016precursors, venkatramani2016precursors}. We attribute the emergence of ordered periodic oscillations from high dimensional chaos to self-organization due to feedback between subsystems.

We explore the scaling behaviour of such self-organization leading to oscillatory instabilities in turbulent fluid mechanical systems. The proximity to the onset of oscillatory instability in each system is quantified using the Hurst exponent (\textit{H}) which also serves as a system independent parameter to study the scaling behaviour of self-organization. An unsteady variable of each of the three systems is measured as we vary an appropriate system-specific control parameter to approach oscillatory instability. We estimate \textit{H}, which is related to the fractal dimension (\textit{D}) as $H = 2 -D$, for the time series corresponding to each state \cite{mandelbrot1982fractal}. 

\section{Results} We analyze the time series of acoustic pressure fluctuations during the transition to oscillatory instabilities for thermoacoustic and aeroacoustic systems. Whereas, in the case of the aeroelastic system, we analyze the time series of strain experienced by the structure. In this work, we study the transition to oscillatory instabilities in the following different cases: (i) a bluff body stabilized combustor of length 700 mm, (ii) and one of length 1400 mm, (iii) a swirl stabilized combustor of length 700 mm, (iv) an aeroacoustic system and (v) an aeroelastic system. We choose these systems as they have different mechanisms for onset of oscillatory instability and have different levels of turbulence, amplitude and frequency of oscillations. The first three cases are for thermoacoustic system wherein the first two cases, the length of the combustor is varied to achieve different acoustic timescales. Similarly, the different flame stabilizing mechanisms renders completely different flow physics. Experimental setups are summarized in Fig.~\ref{fig1} and detailed descriptions of the setups are provided in Appendix B. 

\begin{figure}
\includegraphics[width=0.49\textwidth]{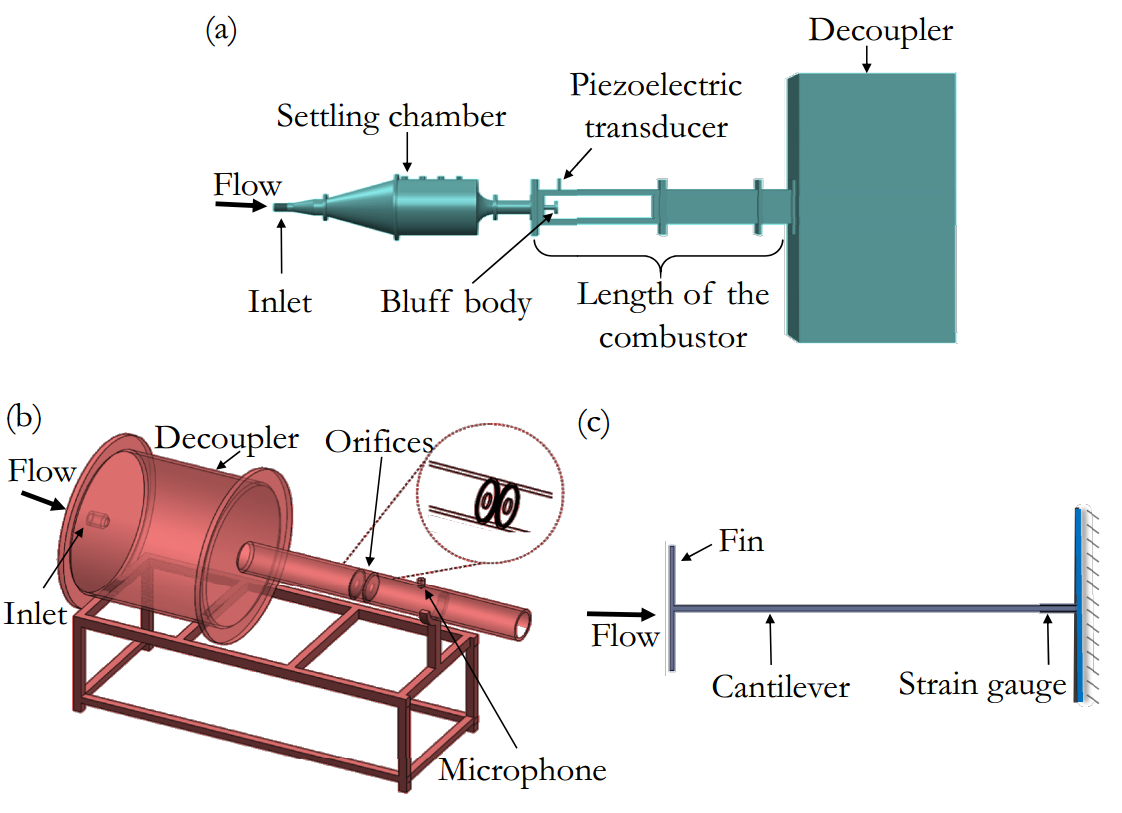}
\caption{Schematic of the experimental setups. (a) Turbulent combustor (thermoacoustic system) exhibiting transition to thermoacoustic instability. Mass flow rate of air is increased by keeping the mass flow rate of fuel constant thus increasing \textit{Re} to attain different dynamical states. The acoustic pressure fluctuations are acquired using a piezoelectric transducer. (b) An aeroacoustic system with two orifices. Vortices are shed when the turbulent flow passes through the orifices. We increase the mass flow rate of the air to achieve different dynamical states. We measure the acoustic pressure fluctuations during the transition to aeroacoustic instability. (c) In the aeroelastic system, the left end of the beam has a small vertical fin attached to it, akin to a winglet of an aircraft wing. When a jet of air passes along the length of the cantilever from left to right, vortices are shed from the fins. These vortices impart an unsteady aerodynamic load to the cantilever. We measure the resulting strain on the cantilever close to the fixed end of the beam. The control parameter in this case is the mean velocity of the jet. }{}{\label{fig1}}
\end{figure}

\begin{figure*}[t]
\includegraphics[width=0.96\textwidth]{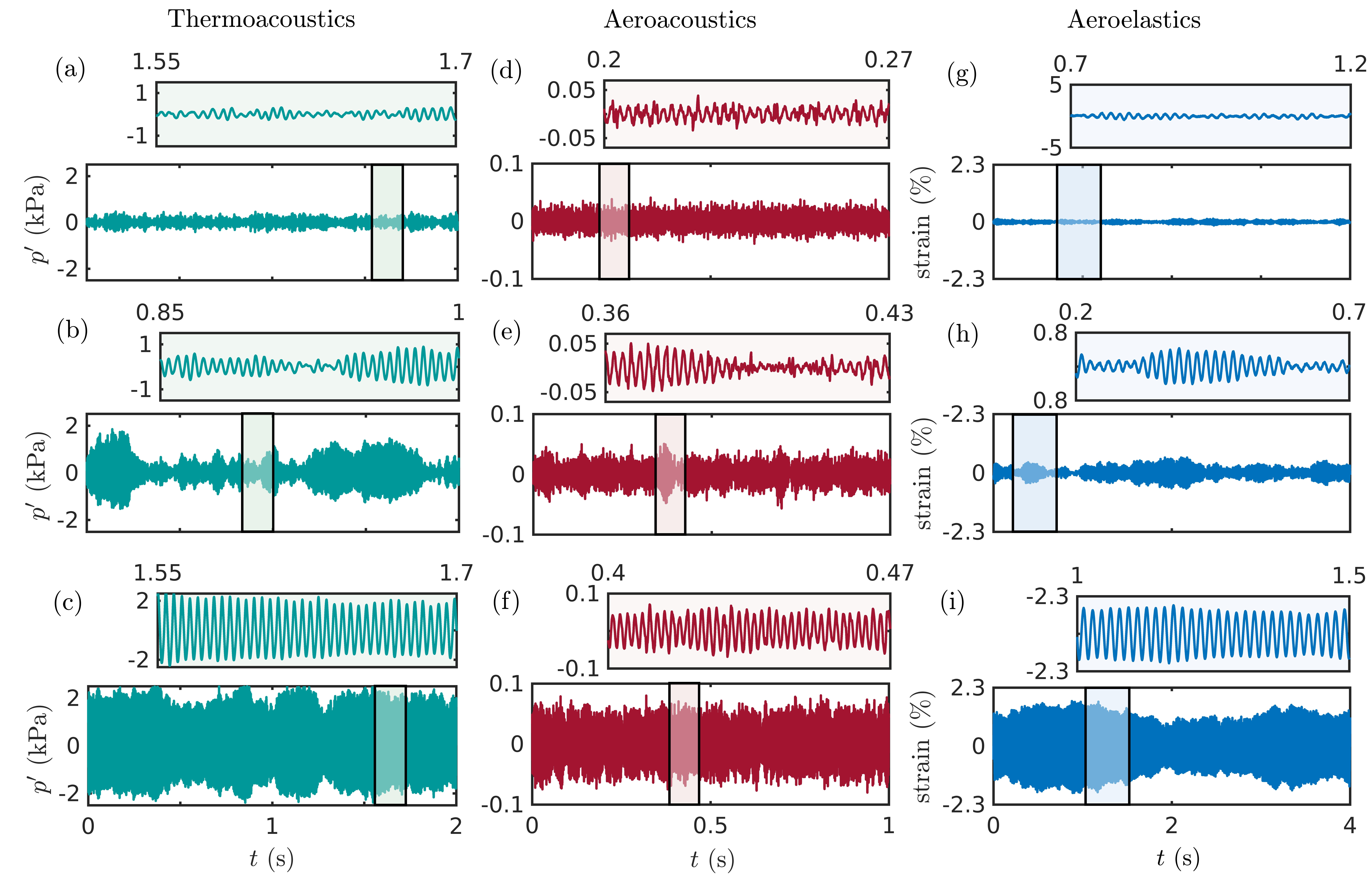}
\caption{\label{fig2} Time series of state variables during the transition to oscillatory instability. (a-c) Data representing the acoustic pressure fluctuations acquired from a bluff body stabilized combustor of length 700 mm. The corresponding \textit{Re} for a, b and c are (1.9 $\pm$ 0.053)x$10^4$, (2.6 $\pm$ 0.069)x$10^4$ and (2.8 $\pm$ 0.073)x$10^4$ respectively. (d-f) Acoustic pressure fluctuations acquired during the transition to aeroacoustic instability (\textit{Re} = 5615 $\pm$ 185, 7283 $\pm$ 198 and 9270 $\pm$ 212 corresponding to d, e and f). (j-i) The time series of strain experienced by the cantilever in the aeroelastic system as we vary \textit{Re} (2384 $\pm$ 111, 3972 $\pm$ 142 and 4768 $\pm$ 159). In all the systems, we observe a transition from low amplitude aperiodic fluctuations (a, d and g) to high amplitude periodic oscillations (c, f and i) via a regime of intermittency where intermittent bursts of high amplitude periodic oscillations appear in a nearly random fashion amidst epochs of low amplitude aperiodic fluctuations (b, e and h) as we vary the control parameters (\textit{Re} increases from top to bottom). The transition from aperiodicity to periodicity always occurs via a regime of intermittency for other configurations of these systems as well.}
\end{figure*}

In Fig.~\ref{fig2}, we show representative datasets from all the three systems. 

\noindent \hangindent=0.4cm I) Figure~\ref{fig2}a-c shows the acoustic pressure fluctuations in a thermoacoustic system (case (i)) during the transition to thermoacoustic instability. Figure~\ref{fig2}a corresponds to a chaotic state far from the oscillatory instability. The time series consists of low amplitude aperiodic fluctuations. Recently, Tony \textit{et al}. \cite{tony2015detecting} showed that these aperiodic fluctuations have features of high-dimensional chaos contaminated with white and coloured noise. Nair \textit{et al}. \cite{nair2014intermittency} discovered that the transition to thermoacoustic instability occurs through a state of intermittency, which contains epochs of high amplitude periodic oscillations amidst low amplitude aperiodic oscillations (Fig.~\ref{fig2}b). Thermoacoustic instability (Fig.~\ref{fig2}c) corresponds to a state of high amplitude periodic oscillations. We observe a similar behaviour for all the above mentioned combustor configurations during the transition to thermoacoustic instability.

\noindent \hangindent=0.4cm II) Figure~\ref{fig2}d-f shows the time series of pressure fluctuations corresponding to the transition to aeroacoustic instability. The temporal behaviour of acoustic pressure during this transition is similar to that in the thermoacoustic system, despite the fact that the amplitude levels in both systems differ by orders of magnitude. 

\noindent \hangindent=0.4cm III) Figure~\ref{fig2}g-i represents the time series of strain experienced by the structure during the transition to aeroelastic instability. The observed oscillations are similar to those of the thermoacoustic and aeroacoustic systems, even though we are measuring a completely different unsteady variable. 

From Fig.~\ref{fig2}, we clearly see that these turbulent systems considered here follow an intermittency route to oscillatory instability. Next, we quantify the proximity to the onset of oscillatory instability in the discussed systems using \textit{H}. As mentioned earlier, the periodic content in time series of the unsteady variable increases as we approach an oscillatory instability. The state of low amplitude aperiodic oscillations has a fractal nature which is born out of the inherent fractal nature of turbulence. As the system self-organizes into oscillatory instability, the fractal time series transitions to a more regular periodic signal \cite{nair2014multifractality}. We capture the variation of fractal characteristics of the time series by calculating \textit{H} following the multifractal detrended fluctuation analysis detailed in the Appendix A.

In Fig.~\ref{fig3}, we plot the amplitude of the dominant mode of oscillations (\textit{A}) and Hurst exponent (\textit{H}) for the time series of pressure oscillations as a function of Reynolds number (\textit{Re}) for the thermoacoustic system (described earlier as case (i)). Note that, $A$ is the amplitude of the dominant peak from the amplitude spectrum of the fluctuating state variable obtained using fast Fourier transform. The signal corresponding to thermoacoustic instability has \textit{H} very close to 0, as the signal is perfectly periodic. We observe that during the transition, \textit{A} increases steeply near the onset of thermoacoustic instability as we vary the control parameter. In contrast, \textit{H} gradually decreases towards zero during the transition. The amplitude of oscillations or the value of \textit{A} at the onset of oscillatory instability depends on the specific system under consideration. In contrast, the variation of \textit{H} describes the self-organization of turbulent flows into oscillatory instabilities, independent of the system features. 

We plot the variation of $A/A_I$ with \textit{H} in log-log scale (Fig.~\ref{fig4}) for the five different cases mentioned earlier. Here, we normalize $A$ of each system with the amplitude of oscillations at the onset of instability ($A_I$) for the given system. We observe that all the data points collapse to a single straight line and this reveals an inverse power law relation between \textit{A} and \textit{H} during the intermittency regime for all the cases considered. Experimentally observed value for the power law exponent is found to remain constant around -2, for all the data irrespective of the frequency of oscillations or the physics of the system. 

Scaling laws and universality are important concepts in statistical physics. They describe the striking similarity in the behaviour during critical transitions among systems that are otherwise different. Scaling in non-equilibrium phase transitions is a topic of interest in recent years \cite{tauber2017phase}. For example, Tham \textit{et al}. \cite{tham1994scaling} experimentally obtained a similar power law scaling relationship between the electrostatic fluctuation levels and the linear growth rate for self-organization in turbulent plasma leading to a quasi-coherent state. 

\begin{figure}
\includegraphics[width=0.49\textwidth]{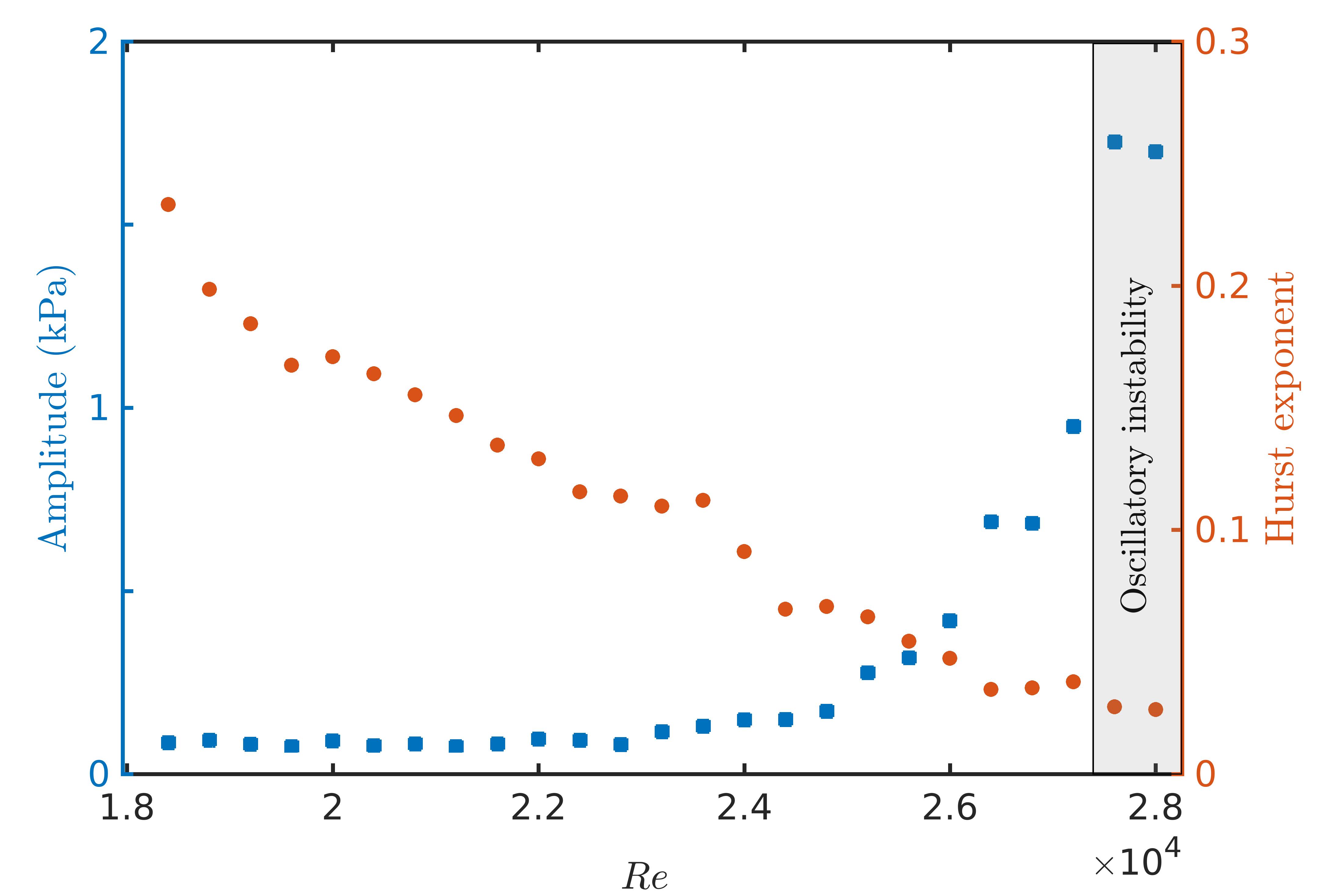}
\caption{\label{fig3} Amplitude of the dominant mode of oscillations and the Hurst exponent for unsteady pressure signals as a function of Reynolds number (\textit{Re}). We analyze the data from a laboratory bluff body stabilized combustor of length 700 mm for different \textit{Re}. The amplitude is obtained from the amplitude spectrum plotted with a resolution of 4 Hz. The amplitude increases steeply near the transition to thermoacoustic instability, whereas the Hurst exponent shows a gradual decrease during the transition and it is approaching zero. 
}
\end{figure}

\section{Discussions} Transition to oscillatory instability in the class of turbulent fluid mechanical systems discussed here occurs via the state of intermittency and we observe a universal scaling law during the transition. In fluid dynamics literature, intermittency refers to a state in which a laminar flow is interrupted by high amplitude turbulent bursts at apparently random intervals \cite{nayfehapplied}. During the bursts, the phase space trajectory goes to a larger chaotic attractor with the original periodic attractor as its subset. Three types of bifurcations are associated with such intermittencies, namely, cyclic fold, subcritical Hopf, and subcritical period-doubling bifurcations. Intermittencies corresponding to these bifurcations are labelled as type I, II and III, respectively \cite{manneville1979intermittency, pomeau1980intermittent}. \footnote{Several other types of intermittencies have been reported and discussed \cite{schuster2006deterministic}.} 

In our case, to begin with, the system is chaotic and is Lyapunov stable. However, during intermittency, this stability is lost and the system intermittently approaches limit cycle oscillations. In contrast to the known types of intermittencies discussed above, here, the intermittency comprises of high amplitude periodic oscillations amidst epochs of low amplitude aperiodic oscillations \cite{pawar2018intermittency}. The trajectory in the phase space goes to a larger periodic attractor from a smaller chaotic attractor during the intermittent bursts (Fig.~\ref{fig2} b, e and h). Thus, there is an inherent difference in the type of intermittency observed during the emergence of oscillatory instabilities in turbulent flows, as observed for example in thermoacoustic, aeroacoustic and aeroelastic systems compared to the classical ones.
\begin{figure}
\includegraphics[width=0.49\textwidth]{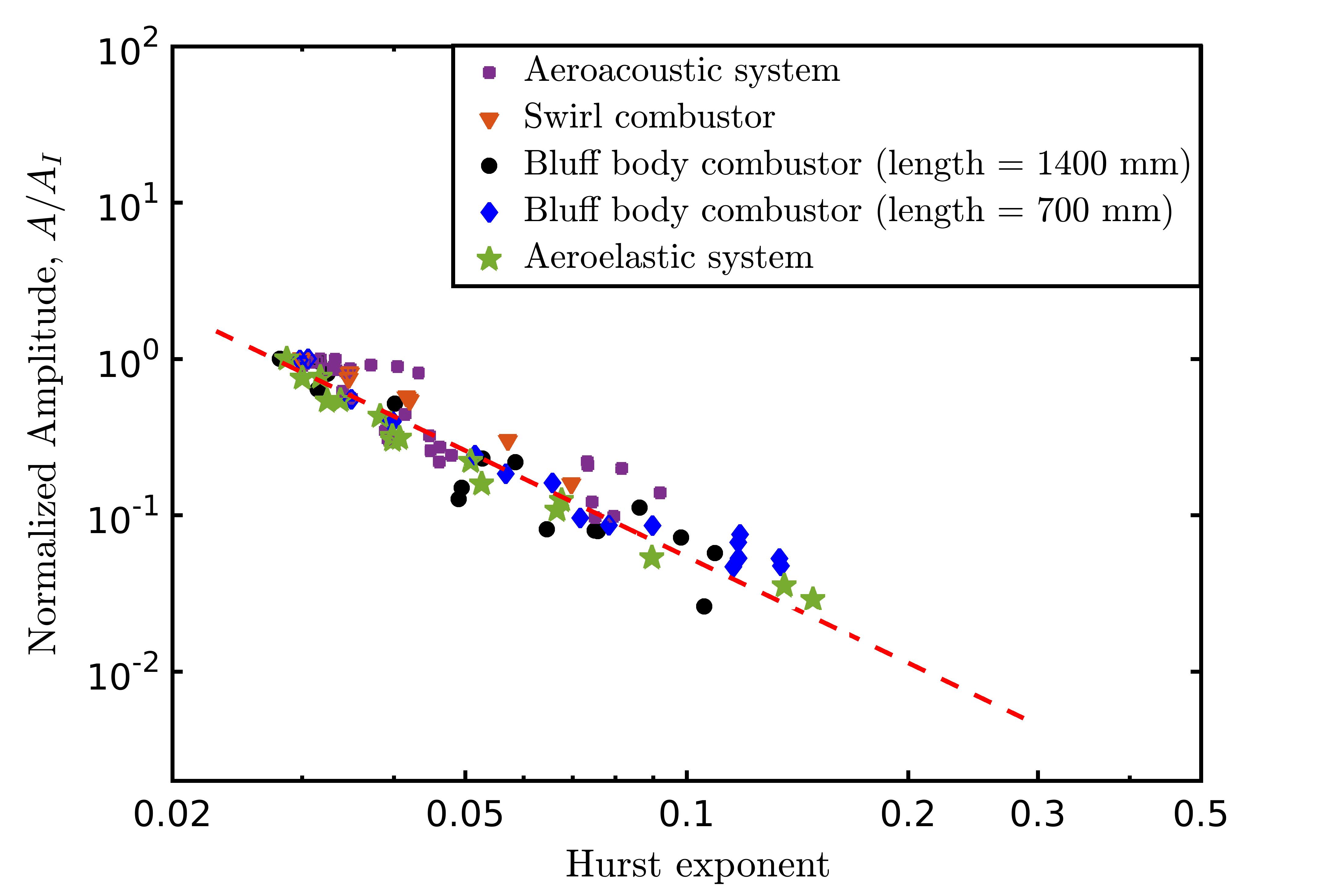}
\caption{\label{fig4} Inverse power law scaling of amplitude with Hurst exponent. Variation of amplitude with Hurst exponent is plotted on a logarithmic scale for the data acquired from different systems. We observe a power law relation with a constant power law exponent around -2 (-1.83 $\pm$ 0.17 for the bluff body combustor with length 700 mm, -2.22 $\pm$ 0.58 for the bluff body combustor with length 1400 mm, -2.06 $\pm$ 0.24 for the swirl combustor, -2.02 $\pm$ 0.32 for the aeroacoustic system and -2.21 $\pm$ 0.19 for the aeroelastic system). The uncertainties are estimated for 90\% confidence intervals. The points with $H > 0.1$ are ignored while finding the power law exponent as they represent the low amplitude aperiodic oscillations far away from the self-organized state.}
\end{figure}

In the present study, we observe the scaling behaviour in all the systems we have examined, where oscillatory instabilities emerge in turbulent flows. We do not observe this inverse power law relation in models such as kicked oscillator \cite{seshadri2016reduced} or noisy Hopf bifurcations \cite{noiray2017linear}, even though they capture the transition from chaos to limit cycle via intermittency. Further, this scaling is not exhibited by models which capture the transition from chaos to periodic oscillations through type I, II and III intermittencies (shown in the supplementary material). This experimentally observed scaling appears like a universal property for a class of systems in which order emerges from chaos, as a result of self-organization in turbulence following an intermittency route. 

Fully developed, isotropic turbulence has a well-known power-law scaling for its energy spectrum \cite{richardson1926atmospheric, kraichnan1967inertial}, which shows the distribution of energy across different wave numbers. The instances of self-organization in turbulence leading to oscillatory instability discussed in this paper are associated with the emergence of periodically shed, large coherent structures in the flow. This is accompanied by the redistribution of energy across different length scales and thus deviation from the scaling observed in fully developed turbulent flows. In the various systems which we examine, as we approach oscillatory instabilities by changing some control parameter of each system, the redistribution of energy into the most dominant scale (\textit{i.e}., scale of coherent structure) in each system is captured by studying the amplitude spectra of an appropriate state variable of the system. In our study, we used unsteady pressure measurements for thermoacoustic and aeroacoustic systems and strain rate for the aeroelastic system. 

\section{Conclusions} In the present study, using three different systems, we describe a universal route through which oscillatory instabilities emerge in turbulent flow. The amplitude of the dominant mode of oscillations increases following an inverse power law scaling with the Hurst exponent of the time series of the appropriate state variable, and the scaling exponent is invariant across the three systems considered. The proximity to the onset of oscillatory instabilities is quantified by the Hurst exponent, which serves as a system independent measure of self-organization. Here, the spectral amplitude of the dominant mode of oscillations serves as the order parameter of the system.

Power law scaling have been discovered for various critical transitions. Here, we report the experimental observation of a scaling behaviour $(A \propto H^{-2})$ for a class of non-equilibrium systems. The discovery of this unique scaling enables a priori estimation of the amplitude of oscillations. This information can be critical in devising the counter measures needed to limit the possible damages from such oscillatory instabilities.

\acknowledgments
We express our gratitude to the Department of Science and Technology, Government of India for providing financial support under the grant numbers: DST/SF/1(EC)/2006 (Swarnajayanti Fellowship) and JCB/2018/000034/SSC (JC Bose Fellowship). We acknowledge the discussions and help from Prof. G. Thampi, Prof. V. Nair, Mr. Manikandan R, Mr. Thilagraj and Mr. Anand. Induja Pavithran is indebted to Ministry of Human Resource Development, India and Indian Institute of Technology Madras for providing research assistantship.

\bibliographystyle{eplbib}
\bibliography{epl}

\providecommand{\noopsort}[1]{}\providecommand{\singleletter}[1]{#1}%
\begin{thebibliography}{10}
\expandafter\ifx\csname url\endcsname\relax\def\url#1{\texttt{#1}}\fi

\bibitem{richardson1926atmospheric}
\Name{Richardson L.~F.} \REVIEW{Proceedings of the Royal Society of London.
  Series A, Containing Papers of a Mathematical and Physical
  Character}{110}{1926}{709}.

\bibitem{kraichnan1967inertial}
\Name{Kraichnan R.~H.} \REVIEW{The Physics of Fluids}{10}{1967}{1417}.

\bibitem{juniper2018sensitivity}
\Name{Juniper M.~P. \and Sujith R.} \REVIEW{Annual Review of Fluid
  Mechanics}{50}{2018}{661}.

\bibitem{flandro2003aeroacoustic}
\Name{Flandro G.~A. \and Majdalani J.} \REVIEW{AIAA journal}{41}{2003}{485}.

\bibitem{hansen2007aeroelastic}
\Name{Hansen M.~H.} \REVIEW{Wind Energy: An International Journal for Progress
  and Applications in Wind Power Conversion Technology}{10}{2007}{551}.

\bibitem{lieuwen2005combustion}
\Name{Lieuwen T.~C. \and Yang V.} \Book{Combustion instabilities in gas turbine
  engines: operational experience, fundamental mechanisms, and modeling}
  (American Institute of Aeronautics and Astronautics) 2005.

\bibitem{fleming1998turbine}
\Name{Fleming C.} \REVIEW{Wall Street Journal}{}{February 13, 1998}{}.

\bibitem{fisher2009remembering}
\Name{Fisher S.~C. \and Rahman S.~A.} \Book{Remembering the giants: Apollo
  rocket propulsion development} (NASA Monographs in Aerospace History series)
  2009.

\bibitem{kriesels1995high}
\Name{Kriesels P., Peters M., Hirschberg A., Wijnands A., Iafrati A., Riccardi
  G., Piva R. \and Bruggeman J.} \REVIEW{Journal of Sound and
  Vibration}{184}{1995}{343}.

\bibitem{larsen1997aeroelastic}
\Name{Larsen A. \and Walther J.~H.} \REVIEW{Journal of Wind Engineering and
  Industrial Aerodynamics}{67}{1997}{253}.

\bibitem{nair2014intermittency}
\Name{Nair V., Thampi G. \and Sujith R.} \REVIEW{Journal of Fluid
  Mechanics}{756}{2014}{470}.

\bibitem{nair2016precursors}
\Name{Nair V. \and Sujith R.} \REVIEW{International Journal of
  Aeroacoustics}{15}{2016}{312}.

\bibitem{venkatramani2016precursors}
\Name{Venkatramani J., Nair V., Sujith R., Gupta S. \and Sarkar S.}
  \REVIEW{Journal of Fluids and Structures}{61}{2016}{376}.

\bibitem{mandelbrot1982fractal}
\Name{Mandelbrot B.~B.} \Book{The fractal geometry of nature} Vol.~1 (WH
  freeman New York) 1982.

\bibitem{tony2015detecting}
\Name{Tony J., Gopalakrishnan E., Sreelekha E. \and Sujith R.} \REVIEW{Physical
  Review E}{92}{2015}{062902}.

\bibitem{nair2014multifractality}
\Name{Nair V. \and Sujith R.} \REVIEW{Journal of Fluid
  Mechanics}{747}{2014}{635}.

\bibitem{tauber2017phase}
\Name{T{\"a}uber U.~C.} \REVIEW{Annual Review of Condensed Matter
  Physics}{8}{2017}{185}.

\bibitem{tham1994scaling}
\Name{Tham P. \and Sen A.} \REVIEW{Physical review letters}{72}{1994}{1020}.

\bibitem{nayfehapplied}
\Name{Nayfeh A.~H. \and Balachandran B.} \REVIEW{Wileylnterscience, New
  York}{}{2008}{}.

\bibitem{manneville1979intermittency}
\Name{Manneville P. \and Pomeau Y.} \REVIEW{Physics Letters A}{75}{1979}{1}.

\bibitem{pomeau1980intermittent}
\Name{Pomeau Y. \and Manneville P.} \REVIEW{Communications in Mathematical
  Physics}{74}{1980}{189}.

\bibitem{schuster2006deterministic}
\Name{Schuster H.~G. \and Just W.} \Book{Deterministic chaos: an introduction}
  (John Wiley \& Sons) 2006.

\bibitem{pawar2018intermittency}
\Name{Pawar S.~A. \and Sujith R.} \Book{Intermittency: A state that precedes
  thermoacoustic instability} in \Book{Droplets and Sprays} (Springer) 2018 pp.
  403--430.

\bibitem{seshadri2016reduced}
\Name{Seshadri A., Nair V. \and Sujith R.} \REVIEW{Combustion Theory and
  Modelling}{20}{2016}{441}.

\bibitem{noiray2017linear}
\Name{Noiray N.} \REVIEW{Journal of Engineering for Gas Turbines and
  Power}{139}{2017}{041503}.

\end{thebibliography}






\end{document}